\documentclass[aps,prb,preprint,showpacs,superscriptaddress]{revtex4}
\usepackage{graphicx}
\usepackage{bm}
\newcommand{\beq}{\begin{equation}}
\newcommand{\eeq}{\end{equation}}
\begin{document}
\title{\boldmath Entropy paradox in strongly correlated Fermi systems}
\author{J.~W.~Clark}
\affiliation{McDonnell Center for the Space Sciences \&
Department of Physics, Washington University,
St.~Louis, MO 63130, USA}
\author{M.~V.~Zverev}
\affiliation{Russian Research Centre Kurchatov
Institute, Moscow, 123182, Russia}
\affiliation{Moscow Institute of Physics and Technology, Moscow, 123098, Russia}
\author{V.~A.~Khodel}
\affiliation{Russian Research Centre Kurchatov
Institute, Moscow, 123182, Russia}
\affiliation{McDonnell Center for the Space Sciences \&
Department of Physics, Washington University,
St.~Louis, MO 63130, USA}

\date{\today}
\begin{abstract}
A system of interacting, identical fermions described by standard Landau
Fermi-liquid (FL) theory can experience a rearrangement of its Fermi surface
if the correlations grow sufficiently strong, as occurs at a quantum critical
point where the effective mass diverges.  As yet, this phenomenon defies
full understanding, but salient aspects of the non-Fermi-liquid (NFL) behavior
observed beyond the quantum critical point are still accessible within the
general framework of the Landau quasiparticle picture.  Self-consistent
solutions of the coupled Landau equations for the quasiparticle momentum
distribution $n(p)$ and quasiparticle energy spectrum $\epsilon(p)$ are
shown to exist in two distinct classes, depending on coupling strength and
on whether the quasiparticle interaction is regular or singular at zero
momentum transfer.  One class of solutions maintains the idempotency condition
$n^2(p) = n(p)$ of standard FL theory at zero temperature $T$ while adding
pockets to the Fermi surface.  The other solutions are characterized by a
swelling of the Fermi surface and a flattening of the spectrum $\epsilon(p)$
over a range of momenta in which the quasiparticle occupancies lie between
0 and 1 even at $T = 0$.  The latter, non-idempotent solution is revealed
by analysis of a Poincar\'e mapping associated with the fundamental Landau
equation connecting $n(p)$ and $\epsilon(p)$ and validated by solution of
a variational condition that yields the symmetry-preserving ground state.
Paradoxically, this extraordinary solution carries the burden of a large
temperature-dependent excess entropy down to very low temperatures, threatening
violation of the Nernst Theorem.  It is argued that certain low-temperature
phase transitions, notably those involving Cooper-pair formation, offer
effective mechanisms for shedding the entropy excess.  Available measurements
in heavy-fermion compounds provide concrete support for such a scenario.
\end{abstract}
\maketitle

\section{Introduction}

Landau Fermi liquid (FL) theory\cite{lan1,lan2,trio,lanl} is universally recognized
as a cornerstone of modern low-temperature condensed matter physics.   This
theory predicts that the magnetic susceptibility $\chi(T)$ becomes independent
of temperature $T$ as $T \to 0$.  It also predicts that in this regime
the entropy $S(T)$ varies linearly with $T$, implying the same behavior
for the specific heat $C(T)=TdS(T)/dT$ and thermal expansion coefficient
$\alpha=l^{-1}\partial l/\partial T \propto -\partial S(T)/\partial P$.
These predictions are in excellent agreement with available experimental
data on conventional Fermi liquids, notably three-dimensional (3D) liquid
$^3$He and the electron liquid present in ordinary metals.

However, beginning the mid-1990's experimental studies of many-fermion systems
have yielded abundant evidence of the failure of the FL picture upon entry
into the regime of strong correlations among the particles.  Such non-Fermi-liquid
(NFL) behavior was first revealed and explored in experiments
\cite{godfrin1995,godfrin1996,godfrin1998,saunders1,saunderssc} on films of
liquid $^3$He and found subsequently in electron systems in solids, especially
heavy-fermion compounds.\cite{loh,steglich,shagrep}  One of the striking features
of the observed NFL behavior is that as the temperature $T$ drops to zero, both the
spin susceptibility $\chi(T)$ and the Sommerfeld ratio
$\gamma(T)=C(T)/T\equiv dS(T)/dT$ diverge when the density
(or doping) reaches a critical value, while the Sommerfeld-Wilson
ratio $R_{SW}=\chi(T)/\gamma(T)$ changes much more slowly.

The corresponding point in the Lifshitz phase diagram showing the sequence
of different phases that replace one another at $T=0$ under variation of a suitable
control parameter (coupling strength, density, doping) is generally called
the quantum critical point (QCP).  If the QCP were the end point, from finite
$T$ values, of a line of transition points $T_c(B)$ associated with some
control parameter $B$, then according to the theory of second-order phase
transitions, divergence of $\gamma(T)$ and $\chi(T)$ at $T\to 0$ would be
specified by different critical indexes.  For example, in the Landau theory
of second-order phase transitions, the critical index $\varsigma$ characterizing
the divergence of the Sommerfeld ratio is zero, while that for the divergence
of the magnetic susceptibility $\chi$ is $4/3$.  In Wilson theory, the critical
indexes have almost the same values as in Landau theory.\cite{wilson}  However, 
experiment deviates strongly from such theoretical predictions.  For example,
the critical-index value $\varsigma = 0.38$ has been determined\cite{krellner}
for the compound YbRh$_2$Si$_2$, which belongs to a family of heavy-fermion
metals comprehensively studied by F.\ Steglich and his
collaborators.\cite{trovar,geg,hossain,oeschler,gegs}

Further, the posited second--order phase transition, attributed as usual
to critical spin fluctuations, possesses unusual properties: the magnetic
moment specifying the ordered state of the metal YbRh$_2$Si$_2$ is extremely
small, $\simeq 10^{-3}\mu_B$, as if the order parameter were hidden.\cite{musr}
Furthermore, in dense $^3$He films, where the emergent NFL behavior was
documented for the first time, experiment has not identified any related
second-order phase transition. Thus the theory of second-order phase
transitions is hardly relevant to explanation of the observed NFL behavior.

Consider, on the other hand, that within FL theory the quantities $\chi$ and
$\gamma$ are both proportional to the density of states $N(0)$.  In turn, $N(0)$
is a linear function of the effective mass $M^*$, which specifies, via the dispersion
$d\epsilon(p\to p_F)/dp=p_F/M^*$, the spectrum $\epsilon(p)$ of single-particle
excitations near the Fermi surface.  Thus, the NFL behavior emerging at the
QCP can be explained within the broader framework of FL theory by attributing
the common divergence of $\chi$ and $\gamma$ to divergence of the
effective mass $M^*$ at the QCP.

The NFL behavior of the entropy $S$, which is the subject of the present
article, is exhibited primarily in the divergence, at the QCP, of the
thermal expansion coefficient $\alpha(T\to 0)$.  The most challenging
data have been obtained for a group of heavy-fermion metals.\cite{oeschler}
For CeCoIn$_5$ in particular, $\alpha(T \to 0) $ is found to be almost independent
of temperature and exceeds typical values for ordinary metals by a huge factor
$10^3$--$10^4$, implying a corresponding enhancement of the entropy itself
(see Fig.~\ref{fig:cecointherm}).

Such anomalous behavior of the entropy $S(T\to 0)$ is documented not only in the
electron liquid within solids but also in liquid $^3$He films.\cite{saunderssc}
The empirical evidence implies the existence of a group of Fermi systems
that exhibit paradoxical behavior of the entropy at the lowest temperatures
currently accessible to measurements, behavior in contradiction to the famous
Nernst theorem requiring $S(T=0)=0$.
Our objective is to elucidate this behavior, which was already predicted
two decades ago in Ref.~\onlinecite{ks} and was recently rediscovered in a
very different theoretical context,\cite{lee} namely the finite-charge-density
sector of conformal field theory (CFT) based on the AdS/CFT gravity/gauge duality.
Briefly stated, the origin of the NFL behavior of the entropy lies in
the {\it flattening} of the single-particle spectrum that occurs near the Fermi
surface when the strength of correlations attains a critical level that
triggers this phenomenon, which may be envisioned as a swelling of the
Fermi surface.

\section{Reprise of the standard FL quasiparticle picture}

To set the stage for discussion of the issue of entropy excess
in strongly correlated Fermi systems, let us recall the elements
of the standard Fermi-liquid quasiparticle picture established
by Landau.\cite{lan1,lan2}  At the heart of this picture is
the postulate that there exists a one-to-one correspondence between
the totality of real, decaying single-particle excitations of the
actual Fermi system and a set of immortal interacting quasiparticles,
whose number is equal to the given number $N$ of real particles.
This condition is written as
\beq
{\rm Tr}\int n({\bf p})\,d\upsilon={N\over V}\equiv \rho,
\label{part}
\eeq
where $n({\bf p})$ is the quasiparticle momentum distribution,
$\rho$ is the density, ${\rm Tr}$ implies summation over spin and
isospin variables, and $d\upsilon=d{\bf p}/(2\pi)^D$ is a volume
element in a momentum space of dimension $D$.  In FL theory, all
thermodynamic quantities, such as the ground state energy $E$, the
entropy $S$ etc., are treated as functionals of the quasiparticle
momentum distribution $n({\bf p})$, the entropy $S$ being given
by the {\it standard combinatorial expression}
\beq
S(n)=-{\rm Tr}\int \left(n({\bf p})\ln n({\bf p})
+(1-n({\bf p}))\ln (1-n({\bf p}))\right)d\upsilon .
\label{entr}
\eeq
In the case of homogeneous matter addressed in our treatment, a conventional
variational procedure based on Eq.~(\ref{entr}) and involving restrictions
that impose conservation of the particle number and energy, leads to the
connection\cite{lan1}
\beq
n(p)=\left[ 1+e^{{\epsilon(p)/T}}\right]^{-1}   ,
\label{dist}
\eeq
between the momentum distribution and the quasiparticle energy
\beq
\epsilon(p) ={\delta E\over \delta n(p) } -\mu ,
\label{spec}
\eeq
measured from the chemical potential $\mu$.
Assuming the quasiparticle spectrum to be $T$-independent, one sees
immediately that at $T\to 0$ the quasiparticle momentum distribution $n(p,T)$
coincides with the Fermi step $n_F(p)=\theta(p_F-p)$.

However, as emphasized by Landau,\cite{lan1} the quasiparticle
spectrum $\epsilon(p)$ depends on $T$, a dependence that happens to be
crucial for resolution of issues addressed in this article.
The structure of the spectrum $\epsilon(p)$ can be derived
from the Landau equation\cite{lan1,lan2,lanl,trio}
\beq
v(p)\equiv {\partial\epsilon(p)\over\partial {\bf p}} =
  {{\bf p}\over M} + \int\! f({\bf p},{\bf p_1})\,
  {\partial n(p_1)\over\partial {\bf p_1}}\, d\upsilon_1.
\label{lansp}
\eeq
This relation expresses the deviation of the quasiparticle group
velocity $v(p)$ from the bare particle velocity $p/M$ (with $M$ the
bare particle mass), in terms of the Landau interaction function
$f({\bf p},{\bf p}_1)$.  The latter function is commonly treated as
a phenomenological input and specified by a set of phenomenological
parameters associated with the harmonics of its Legendre polynomial
expansion.  The three relations (\ref{part}), (\ref{dist}), and
(\ref{lansp}) then determine the momentum distribution, quasiparticle
spectrum, and chemical potential self-consistently.

Half a century ago when FL theory was created, solutions of such a complicated
integro-differential equation as (\ref{lansp}) were unavailable. Fortunately,
in ordinary Fermi liquids such as bulk liquid $^3$He or the conduction electron
system in familiar metals, correlations between particles are moderate, allowing
thermodynamic properties of these systems to be adequately described without
the aid of a computer.  Indeed, in these systems, the Fermi velocity
$v_F=v(p_F)\equiv p_F/M^*$ maintains a positive value, implying that the
effective mass $M^*$ is the single parameter that the characterizes solution
\beq
\epsilon(p)=p_F(p-p_F)/M^*
\label{efl}
\eeq
of Eq.~(\ref{lansp}) in the domain $|\epsilon(p)|\leq T$ relevant to the
thermodynamics of Fermi systems.

According to Eq.(\ref{lansp}), the quasiparticle effective mass of a 3D Fermi
system is expressed as follows\cite{trio,lanl}
\beq
{M\over M^*}=1-{1\over 3}F^0_1\equiv
1- f_1{p_FM\over 3\pi^2}
\label{mqcp}
\eeq
in terms of the first harmonic $f_1$ of the interaction function $f({\bf p},{\bf p}_1)$,
where the dimensionless parameter $F^0_1=f_1p_FM/\pi^2$ has been introduced.
Since the spectrum (\ref{efl}) differs from that of an ideal Fermi gas merely
by the numerical factor $M^*/M$, the low-temperature thermodynamic properties
of conventional Fermi liquids---for which $M^*$ remains finite---behave
in just the same way as those of the ideal-gas systems apart
from trivial scaling.  In particular, the spin susceptibility $\chi(T)$
and Sommerfeld ratio $\gamma(T)=C(T)/T$ differ from their results
for the corresponding ideal-gas system simply by the factor $M^*/M$.
Furthermore, upon inserting Eq.~(\ref{efl}) into Eq.~(\ref{dist}) we find
that the $T=0$ quasiparticle momentum distribution $n(p)$ coincides
with the momentum distribution $n_F(p)=\theta(p_F-p)$ of the ideal Fermi gas.
Referring to Eq.~(\ref{entr}), it is seen that the entropy of the
interacting system satisfies $S(T=0)$ in obedience to the Nernst theorem.

\section{Divergence of the effective mass at the QCP}

This time-honored FL quasiparticle picture, in which a real system of
interacting fermions is treated as a ``gas of interacting
quasiparticles,''\cite{migt} worked flawlessly for over three decades,
but as pointed out in the introduction, more recent experimental studies,
especially of novel materials, have brought indisputable evidence for
its failure when the correlations grow too strong.  Within FL theory, the
strength of correlations is measured by the magnitudes of leading
harmonics of the dimensionless interaction function
$F=fN(0)$ where $N(0)=p_FM^*/\pi^2$ is the density of states.  The Fermi
system is considered to be strongly correlated when one (or both) of the
leading Landau parameters $F_0$ or $F_1$ reaches a value close to 1.
Since the effective mass diverges at the QCP, both these parameters diverge as
well, so the QCP regime does indeed belong to a region of the Lifshitz phase
diagram occupied by strongly correlated Fermi systems.  The proportionality to
$N(0)$ of both the spin susceptibility $\chi(T)$ and the Sommerfeld ratio
$\gamma(T)$ implies that they become divergent as the temperature $T$ goes
to zero and hence exhibit NFL behavior.  At the same time, the QCP Sommerfeld-Wilson
ratio $R_{SW}=\chi(T)/\gamma(T)$ remains finite.  Additionally, it should be
noted that the divergence of the specific heat at the QCP means that the
entropy $S(T)$ must also begin to participate in NFL behavior.

The emergence of the QCP is a conspicuous signature of the essential qualitative
difference between Fermi systems that are strongly correlated and those with
weak or moderate correlations.  Three-dimensional (3D) liquid $^3$He, for which
$M^*$ remains finite at any density, belongs to the class of systems
with moderate correlations.  On the other hand, its 2D counterpart
has been assigned to the class of strongly correlated Fermi systems
based on evidence for the existence of a QCP from experimental
studies of dense $^3$He films, as cited above.

The quest for a fundamental understanding of QCP phenomena persists as one
of the most important and most controversial activities in low-temperature
condensed matter physics.  Here we refrain from repeated disputation\cite{prb2008,cmt33}
with respect to numerous papers (see, for example, Ref.~\onlinecite{coleman})
based on the idea that the FL quasiparticle picture (in its wider scope) ceases
to be applicable due to vanishing of the quasiparticle weight $z$ in the
single-particle state, inherent at points of second-order phase transitions.
Aside from such theoretical considerations, this proposal is unable to
provide a consistent description of experimental data in the QCP region.
Rather, we pursue a different approach developed within the broader
framework of the original Landau quasiparticle picture, dating back
to Refs.~\onlinecite{ks,vol,noz,physrep}, explored in detail in
Refs.~\onlinecite{zb,zkb,shaghf,prb2005,yak,prb2008,mig100,shagprb},
sketched above and encapsulated in Eqs.~(\ref{part}), (\ref{dist}),
and (\ref{lansp}).  The emergent theory behaves properly on both the
sides of the QCP,\cite{shagrep,mig100,shagprb} facilitating a consistent
account of the properties of strongly correlated Fermi systems, including
the entropy paradox.

\section{Poincar\'e mapping in the theory of strongly correlated Fermi systems}
The development of a quantitative theory of strongly correlated Fermi systems
within the quasiparticle framework rests upon solution of
the fundamental equation (\ref{lansp}) of FL theory, applied in the
general context where $n(k,T=0)$ need not be the Fermi step
$n_F(p)=\theta(p_F-p)$.   Iteration procedures provide a common strategy
for finding solutions of such nonlinear integro-differential equations,
and we will see that they also pave the way to a resolution of the
entropy paradox that has arisen.  Indeed, the iterative operation
introduced in nonlinear dynamics known as Poincar\'e mapping, and familiar
in the theory of turbulence,\cite{lanh,feig} will prove instrumental to
elucidation of the striking features inherent in solutions of Eq.~(\ref{lansp})
beyond the QCP.

To be more specific, the discrete iterative map corresponding to Eq.~(\ref{lansp})
reads
\beq
{\partial\epsilon^{(j+1)}(p)\over\partial p} ={p\over M} +{1\over
3\pi^2} \int f_1( p, p_1){\partial n^{(j)}(p_1)\over\partial
p_1}p^2_1dp_1  ,
\label{lanit}
\eeq
the iterate $\mu^{(j+1)}$ for the chemical potential being
determined from the normalization condition (\ref{part}).
The index $j = 0,1,2, \ldots$ counts the iterations (zeroth, first,
second, $\ldots$).  The iterate $n^{(j+1)}(p)$ of the momentum
distribution $n(p)$ is generated by inserting the corresponding
spectral iterate $\epsilon^{(j+1)}(p)$ into the right side of Eq.~(\ref{dist}),
which at $T=0$ reduces to a Heaviside function $n(p)=\theta(-\epsilon(p))$.

In canonical Fermi liquids, for which the sign of the Fermi velocity
$v_F$ determined by Eq.~(\ref{mqcp}) is positive and the function
$\epsilon^{(1)}(p)$ has the single zero at $p=p_F$, the first iterate
$n^{(1)}(p)$ and all higher iterates for the distribution
$n(p)$ coincide with $n_F(p)= \theta(p_F -p)$, this being a fixed
point of the transformation.  However, beyond the QCP the sign of
the group velocity $v_F$ becomes negative, and the first iterate
$\epsilon^{(1)}(p)\equiv \epsilon(p;n_F)$ for the spectrum already
has two or three zeroes, implying the emergence of new pockets of
the Fermi surface, or equivalently, new kinks
in the momentum distribution $n^{(1)}(p)$.

What happens in higher iterations depends on the structure of the
interaction function $f$.  Usually this function is smooth, i.e. regular,
in momentum space, although the {\it bare} interaction may have long-range
character in coordinate space.  In this case, transitions occurring beyond the
QCP were first uncovered theoretically in Ref.~\onlinecite{zb} and later
addressed in Ref.~\onlinecite{shagp}.  An important property of all these
solutions is that they have a multi-connected Fermi surface satisfying
the relation
\beq
n^2(p)=n(p),
\label{imp}
\eeq
which is satisfied trivially by $n(p) = \theta(p_F-p)$ in standard FL theory,
continues to hold, implying that $S(T=0)=0$ in agreement with the Nernst theorem.
Thus, the QCP heralds a sequence of topological rearrangements of the Landau
state in the strong correlation regime through an avalanche of new
pockets of the Fermi surface.

However, this scenario based on a proliferation of new pockets
is not universal.  There exists a different scenario for topological
rearrangement of the Landau state,\cite{ks} which was suggested even
earlier: at some finite critical value of the coupling constant, the
number of bubbles or ``Fermi pockets'' becomes {\it infinite}.  The next
subsection will demonstrate how this more radical scenario emerges
within a suitable iterative procedure.  (Additional details may be found
in Refs.~\onlinecite{prb2008,qgpyaf}.) It is this scenario that
is directly relevant to explanation of the entropy paradox.

\subsection{2-cycles in Poincar\'e mapping for systems with a singular
interaction function $f$}

A remarkable feature of equations considered in modeling turbulence is
the doubling of periods of motion in route to dynamical chaos.\cite{feig,lanh}
The nonlinear system corresponding to Eq.~(\ref{lanit}) can be analyzed
within this context by associating an iteration step with a step in time.
One's first instinct is to object that such a phenomenon should not
occur when Poincar\'e mapping is implemented, and assert that
iterations must converge, since chaos in the classical sense
cannot play a role in the ground states of Fermi liquids at $T=0$,
which are assumed to possess a unique, nondegenerate structure.
And indeed, this general assertion seems to be validated by comments made
in the preceding subsection.  However, this is no longer the case
if the effective interaction $f$ turns out to have long range in coordinate
space, so as to become singular at the origin in momentum space.

For example, the singularity that exists in the transverse bare interaction
\beq
\Gamma^0({\bf p}_1,{\bf p}_2,{\bf q},\omega=0)
=-g {{\bf p}_1\cdot{\bf p}_2-({\bf p}_1\cdot{\bf q})
({\bf p}_2\cdot{\bf q})/q^2 \over q^2}
\label{cur}
\eeq
between quarks in dense quark-gluon plasma is not regularized by polarization
effects, in contrast to the well-known regularization (screening) of
the longitudinal Coulomb interaction due to gauge invariance.
\begin{figure}[t]
\includegraphics[width=1.0\linewidth,height=1.1\linewidth]{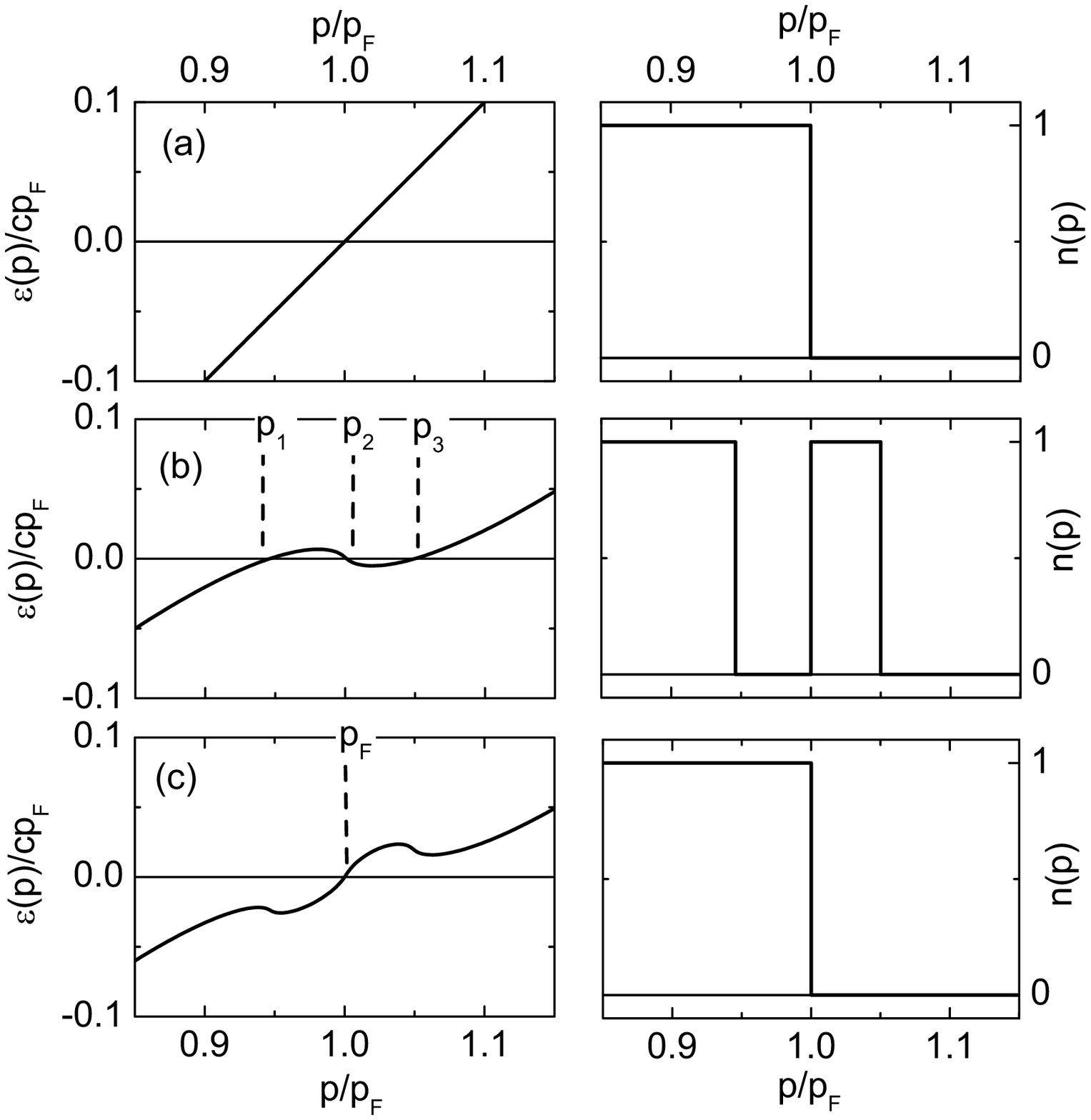}
\caption {Iterative maps for the quark-gluon plasma problem
(\ref{lqgp}) with bare spectrum $\epsilon_p^0=cp$ and
dimensionless coupling parameter $\alpha=g/c=0.3$. Left panels: spectral
iterates $\epsilon^{(j)}(p)$ with $j=0,1,2$, in units of $cp_F$.
Right panels: momentum-distribution iterates $n^{(j)}(p)$.
The 2-cycle reveals itself in the coincidence between the first
and third panels on the right.} \label{fig:fc_cc_2c}
\end{figure}
In this example, Eq.~(\ref{lansp}) takes the form\cite{baym}
\beq
{\partial\epsilon(p)\over \partial p} ={\partial\epsilon^0_p\over
\partial p} +g \int \ln {2p_F\over p-p_1}
{\partial n(p_1)\over \partial p_1} dp_1  ,
\label{lqgp}
\eeq
where $\epsilon^0_p\simeq cp$ is the bare single-particle spectrum
of light quarks, with $c$ the velocity of light.

The first iterate for the spectrum, evaluated from Eq.~(\ref{lqgp}) with
the distribution $n_F(p)$, has an infinite negative derivative
$d\epsilon^{(1)}/dp$ at the Fermi surface.  The corresponding first
iterate of the momentum distribution is $n^{(1)}(p)=\theta(x+x_0)-\theta(x)
+\theta(x-x_0)$, where $x=p/p_F-1$.  The next iteration step yields $n^{(2)}(p)
\equiv n_F(p)$, so the standard FL structure of the momentum
distribution is recovered.  The nonlinear system enters a 2-cycle
that is repeated indefinitely.

The first two iterations of the mapping process are depicted in
Fig.~\ref{fig:fc_cc_2c}. The top-left panel of this figure shows
the bare spectrum $\epsilon^{(0)}(p)\equiv\epsilon^{0}(p)$.  The
first iterate of the spectrum, $\epsilon^{(1)}(p)$, appearing in
the middle-left panel, is evaluated by folding the kernel
$\ln(2p_F/(p-p_1))$ with the Fermi-step $n_F(p)$, shown in the
top-right panel. The spectrum $\epsilon^{(1)}(p)$ possesses three
zeroes: $p_1<p_F$, $p_2=p_F$, and $p_3>p_F$, implying that the
first iterate $n^{(1)}(p)$, drawn in middle-right panel, describes
a Fermi surface with three sheets.  This distribution differs from
the ordinary Fermi step {\it only} in the momentum interval
$-x_0<x<x_0$.  The next iterate, $\epsilon^{(2)}(p)$ (bottom-left
panel), again has a single zero $p_F$, and the corresponding
momentum distribution $n^{(2)}(p)$ (bottom-right panel) coincides
identically with $n_F(p)$.

\begin{figure}[t]
\includegraphics[width=1.0\linewidth,height=1.1\linewidth]{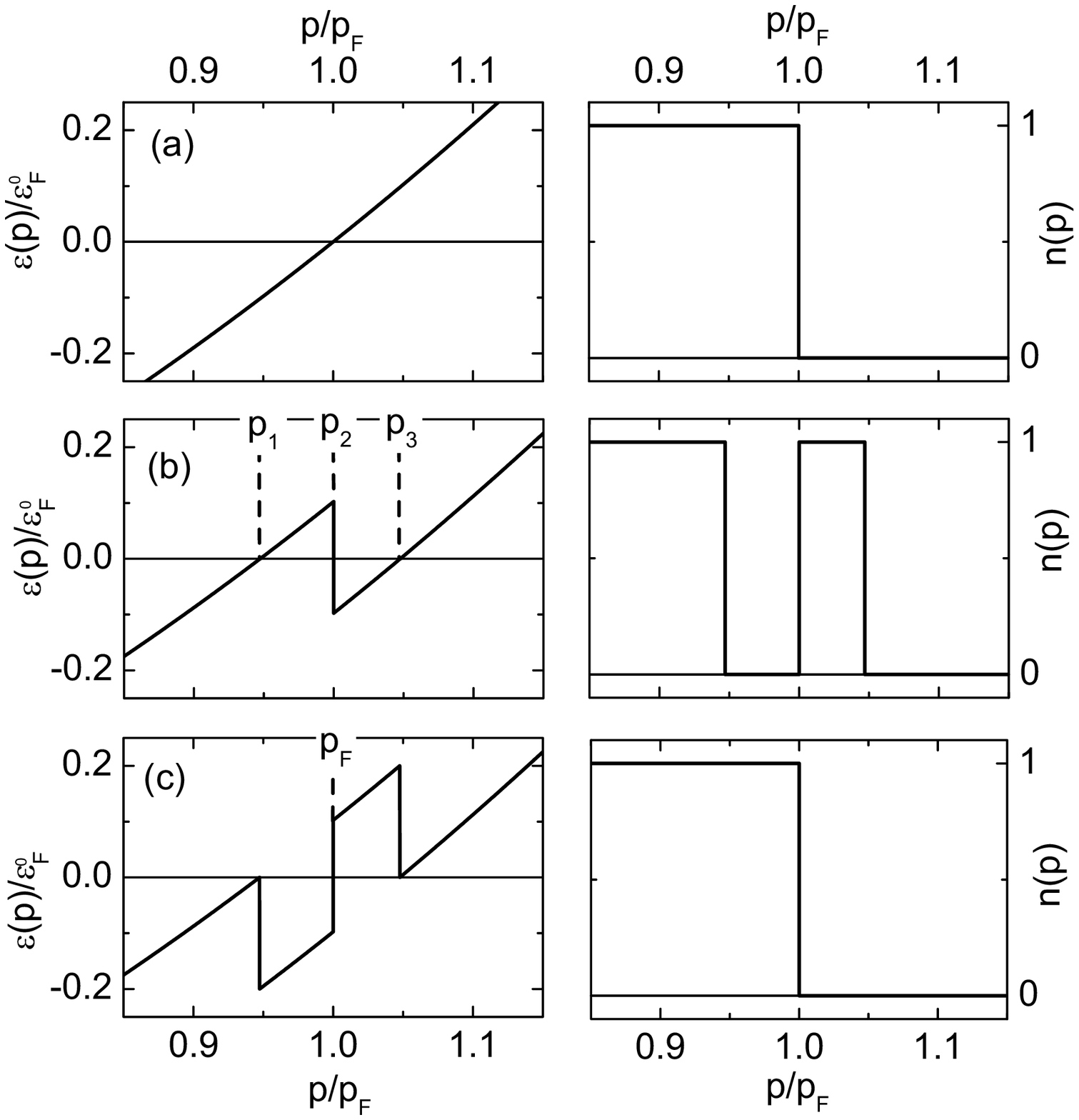}
\caption {Same as in Fig.~\ref{fig:fc_cc_2c} for the discretized
Nozi\`eres model (\ref{nozi}) with $g=0.2\varepsilon_F^0$.}
\label{fig:fc_n_2c}
\end{figure}

This example is not unique in exhibiting 2-cycle terminal behavior.
A 2-cycle Poincar\'e mapping also arises in treating the well-known
Nozi\`eres model,\cite{noz} for which the interaction function
$f$ has the limited singular form $f({\bf q})=(2\pi)^3 g\delta({\bf q})$
with $g>0$.  In this model, the iterative maps (depicted in
Fig.~\ref{fig:fc_n_2c}) are generated from the equation
\beq
\epsilon^{(j+1)}(p)+\mu^{(j+1)}=p^2/2M+gn^{(j)}(p) ,
\label{nozi}
\eeq
along with the normalization condition (\ref{part}) for
$n^{(j+1)}(p)=\theta(-\epsilon^{(j+1)}(p))$.  Here, the odd iterates
$n^{(2j+1)}(p)$ of the momentum distribution deviate from the
$n_F(p)$ in the interval $-g/4\varepsilon^0_F<x<g/4\varepsilon^0_F$,
but in even iterations, the Fermi step reappears intact.

Numerical analysis demonstrates that similar 2-cycles arise when Poincar\'e
mapping based on Eq.~(\ref{lanit}) is implemented for other systems
possessing an interaction function singular at $k\to 0$.  In all
these cases, the emergence of a 2-cycle provides an unambiguous
signal of the instability of the standard Landau state and all its discrete
modifications involving any finite number of pockets of the Fermi
surface.

\subsection{Modified Poincar\'e mapping and new insight from
chaos theory}

Apparently, the occurrence of persistent 2-cycles in the iterative maps
of Eq.~(\ref{lanit}) prevents us from finding self-consistent solutions of
Eq.~(\ref{lansp}) beyond the QCP for a specific class of Fermi systems
possessing singular effective interactions.  It can be argued, however,
that this failure is a consequence of the inadequacy of the iterative
procedure employed, which works perfectly on the FL side of the QCP.
Indeed, a refined procedure that mixes iterations does allow one to
avoid the 2-cycle terminal behavior.  Nevertheless, the improved procedure
still fails to yield a solution: once again the iterations do not
converge,\cite{baym} although the pattern of their evolution becomes
more complicated and---as will be seen---both intriguing and suggestive.

\begin{figure}[t]
\includegraphics[width=.86\linewidth,height=1.21\linewidth]{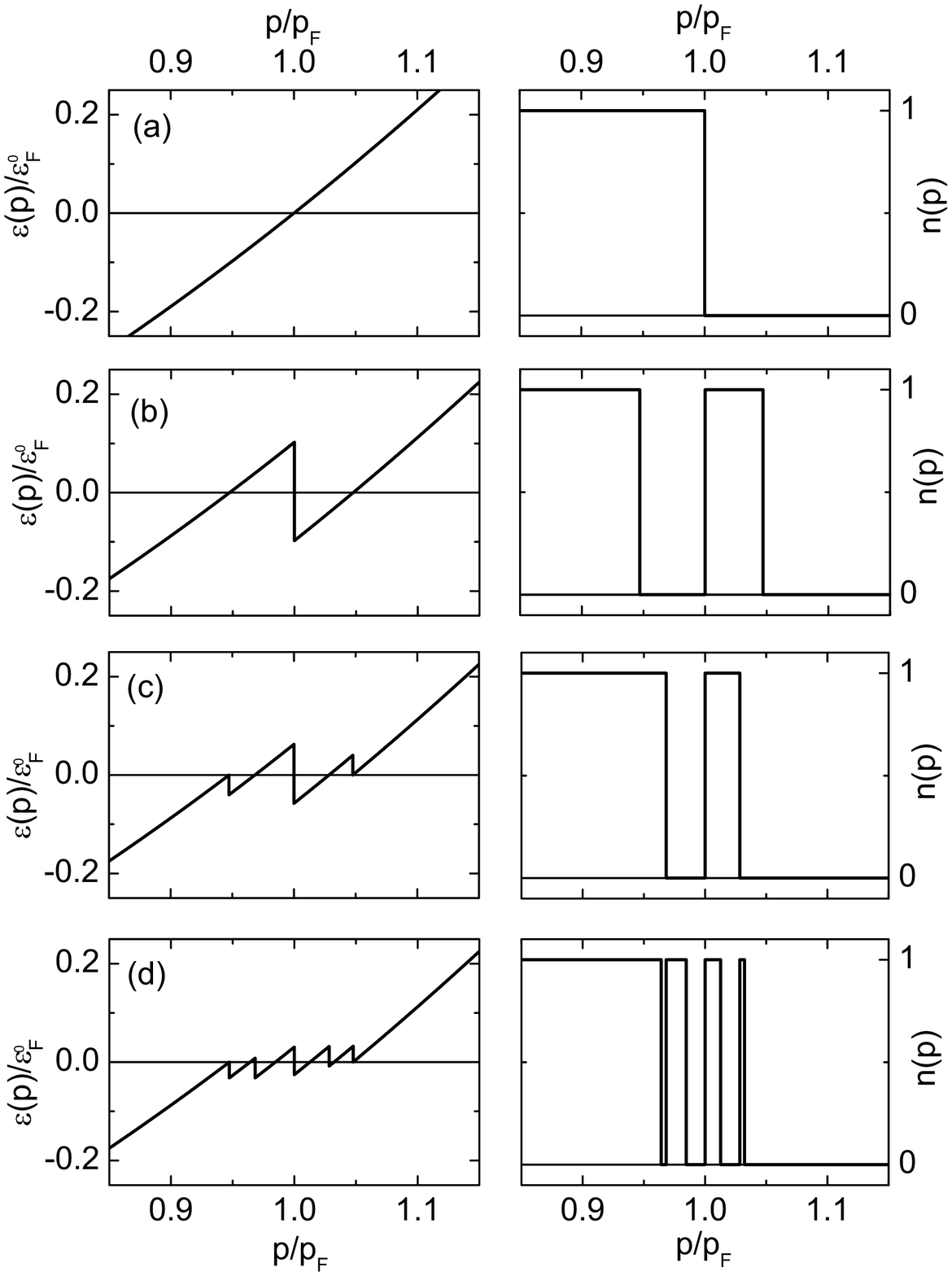}
\caption {Iterative maps for the discretized Nozi\`eres model (\ref{nozi1})
when inputs taken from the two preceding iterations are mixed with
the parameter $\zeta=0.2$.  Left panels: spectra $\epsilon^{(j)}(p)$
with $j=0,1,2,3$, in units of $\varepsilon^0_F$. Right panels:
momentum distributions $n^{(j)}(p)$.
} \label{fig:fc_n_mix}
\end{figure}

By way of illustration, let us consider a refined Poincar\'e mapping for
the Nozi\`eres model, with the same coupling parameter $g=0.2\varepsilon^0_F$
as before.  Choosing a mixing parameter $\zeta$, the equation
\begin{eqnarray}
\lefteqn{\epsilon^{(j+1)}(p)+\mu^{(j+1)}=
(1-\zeta)\,[\epsilon^{(j)}(p) +\mu^{(j)}]} \qquad\qquad
\nonumber \\
& &+\zeta\,[p^2/2M+gn^{(j)}(p)] \qquad\qquad
\label{nozi1}
\end{eqnarray}
is used to generate the iterative maps shown in Fig.~\ref{fig:fc_n_mix}.
Recovery of the ordinary Fermi distribution $n_F(p)$, which is
an inherent feature of the standard iteration procedure at
the each even iteration (see Fig.~\ref{fig:fc_n_2c}),
no longer occurs; indeed, the sequence of iterations fails
to converge, even to a limit cycle.  The number of sheets
remains three at the second iteration.  At the third,
however, seven sheets of the Fermi surface emerge, and
the number of sheets continues to increase in successive
iterations.  The same occurs in dealing with dense quark-gluon plasma.
It is the nonconvergence of iterative procedures that compelled the authors of
Ref.~\onlinecite{baym} to stop searching for solutions of Eq.~(\ref{lqgp}).

Nevertheless, this obstacle can be overcome.  A key to arriving at an acceptable,
self-consistent solution lies in the opportunity  to treat the number
$j=1,2,3,\ldots$ of the iteration as a discrete time step.\cite{prb2008}
In this sense, the sequence of panels in the left column of Fig.~\ref{fig:fc_n_mix}
shows the ``temporal'' evolution of the quasiparticle spectrum.  At any
time-step $t$, the single-particle energy $\epsilon(p,t)$ drops off steadily
as $t$ goes to infinity, while the sign of $\epsilon(p)$ switches unpredictably
in a finite region $\Omega$ of momentum space adjacent to the Fermi surface.
These erratic changes of sign, depending in practice on fluctuations of
voltage, temperature, humidity, etc., that affect the computer executing the
iteration code, induce unpredictable indeterministic jumps of the occupation
numbers $n(p,t)$ between the two values 0 and 1.  On the other hand,
the lack of convergence of the iteration process may be attributed to the
presence of a kind of ``quantum chaos" associated with the basic equation
of FL theory for a certain class of interacting many-fermion systems.  A deeper
understanding of the chaotic evolution of the corresponding systems in the
fictitious time $t$ may well yield unique insights into the role of chaos
theory in the description of quantum phenomena.

Significantly,
as $t$ goes to infinity, the region of momentum space in which iterations of
Eq.~(\ref{lansp}) fail to converge tends to a definite limit $\Omega$.
Entropy is an expected byproduct of chaos.  Accordingly, we tentatively
adopt $\Omega$ as a special entropy that is associated with a Fermi system
under conditions in which iteration of Eq.~(\ref{lansp}) does not converge
to a solution.  This naive {\it Ansatz} for the special entropy can be
refined as follows.  A ``time-averaged'' single-particle energy
${\overline{\epsilon}}(p)$ is defined by the standard formula drawn from
statistical physics,
\beq
{\overline{\epsilon}}(p)=\lim\limits_{T\to \infty}
{1\over T}\int\limits_0^T\epsilon(p,t) dt\equiv
\lim\limits_{J\to\infty} {1\over J}\sum\limits_{j=0}^J\epsilon^{(j)}(p),
\label{epsi}
\eeq
and a corresponding time average
\beq
{\overline{n}}(p)=\lim\limits_{T\to \infty}
{1\over T}\int\limits_0^T n(p,t) dt\equiv
\lim\limits_{J\to\infty} {1\over J}\sum\limits_{j=0}^K n^{(j)}(p)
\label{nav}
\eeq
is constructed for the momentum distribution $n(p)$.  Invoking
Eq.~(\ref{nozi}), the relation between the two time averages is given by
\beq
{\overline{\epsilon}}(p)=p^2/2M -\mu +g {\overline{n}}(p).
\label{Nspav}
\eeq
The mixing parameter $\zeta$ appearing in Eq.~(\ref{nozi1}) cancels out
in deriving this formula.

Wherever the iterations converge, the average ${\overline{n}}(p)$
only takes values 0 or 1, and then ${\overline{\epsilon}}(p)$
is a parabolic function of $p$ coinciding with the true single-particle
energy.  On the other hand, as seen from Eq.~(\ref{epsi}) and verified
by results shown in the left panels of Fig.~\ref{fig:fc_n_mix}, the function
${\overline{\epsilon}}(p)$ {\it vanishes identically} in the domain
$\Omega$ where iterations of Eq.~(\ref{lansp}) do not converge, so that
\beq
n(p) = n_*(p)= {\mu -p^2/2M\over g}\  , \quad  p\in \Omega   .
\label{Nnav}
\eeq
The symbol $n_*(p)$ has been introduced to denote a smoothed momentum
distribution, determined by averaging iterates $\epsilon(p,t)$ for the
single-particle energy according to the prescription (\ref{epsi}).

In summary, by executing (i) a modified Poincar\'e mapping procedure based
on Eq.~(\ref{nozi1}) along with (ii) ``time-averaging'' of iterates for
$\epsilon(p)$ and $n(p)$ in the manner of Eqs.~(\ref{epsi}) and (\ref{nav}),
a self-consistent solution of Eqs.~(\ref{part}), (\ref{dist}), and (\ref{lansp})
that obeys the Pauli principle can in fact be found in the domain $\Omega$ where
the sequence of iterates fails to converge.  The solution so obtained is
independent of the parameters specifying the refined iteration procedure.  The
boundaries $p_i$ and $p_f$ of the momentum interval $p_i<p_F<p_f$ defining
the domain of non-convergence are determined by the conditions $n_*(p_i)=1$
and $n_*(p_f)=0$.  The quasiparticle momentum distribution $n(p)$ corresponding
to the new solution, hereafter written as $n_*(p)$, is given by 1 and 0,
respectively, at $p \leq p_i$ and $p \geq p_f$, and by Eq.~(\ref{Nnav}) in
between.

This program is implemented similarly in investigations of other Fermi
systems for which iteration of Eq.~(\ref{lansp}) does not converge
to a solution.  The central quantity of the iterative procedure, namely the
averaged single-particle energy ${\overline{\epsilon}}(p)$, is constructed
by means of Eq.~(\ref{epsi}).  Since this quantity vanishes identically
in the domain $\Omega$ where iterative solution founders, we immediately
obtain the closed equation
\beq
{\overline{\epsilon}}(p,n_*)=0,  \quad  p \in \Omega
\label {topi}
\eeq
for the smoothed, NFL component of the momentum distribution $n_*(p)$.
Recalling the definition (\ref{spec}) of the quasiparticle energy
$\epsilon(p)$, Eq.~(\ref{topi}) can be recast in the variational form\cite{ks}
\beq
{\delta E(n)\over \delta n(p)}=\mu,
\label{varen}
\eeq
which is free of any trace of  ``time-averaging.''

We can quantitatively characterize the excess entropy $S_*$ of the
presumed ground-state solution at $T=0$
by inserting the momentum distribution $n_*(p)$ determined from Eq.~(\ref{varen})
into the combinatorial expression (\ref{entr}).  By the definition of $n_*(p)$,
the integrand in Eq.~(\ref{entr}) vanishes outside the domain $\Omega$, so
$S_*$ must be essentially proportional to the volume of this domain
(also called $\Omega$).  Setting $n_*(p)$ inside $\Omega$ equal to the
typical value 1/2, one does indeed arrive at $S_*=2\Omega \ln2$.
The existence of such a nonzero entropy excess means that the
rearranged $T=0$ ground state of the system corresponding to this solution
of Eq.~(\ref{lansp}) is {\it statistically degenerate}. In the next
section, it will be demonstrated that such a degeneracy inevitably emerges
at some critical coupling strength in some region of the Lifshitz
phase diagram for {\it any} type of interaction function $f$, whether or
not the effective forces between quasiparticles are of long range.
These findings may open a new thread in discussion of the elusive
concept of ``quantum chaos,'' which has often reduced merely to analysis
of putative chaotic signatures in the spectrum of solutions of quantum
one-body problems.


\section{Flattening of the single-particle spectrum as the source of excess entropy}

As seen above, the NFL results (e.g., Eqs.~(\ref{Nspav}) and (\ref{Nnav})) obtained from
application of the refined iteration procedure for solving the basic FL equation
(\ref{lansp}) can be interpreted as coming from solution of the variational equation for
minimizing the Landau energy functional $E(n)$, with the chemical potential $\mu$
serving as a Langrange multiplier to conserve particle number.

\subsection{Breakdown of the necessary stability condition for the Landau
state and the emergence of flat bands}

We  may now describe, within in a unified theoretical framework, how two alternative
and qualitatively distinct topological transformations of the Fermi surface
can take place when conventional FL theory, characterized by the zero-temperature
Fermi-gas quasiparticle momentum distribution $n_F(p)=\theta(p_F-p)$, gives way to a
more general theory of correlated Fermi systems that maintains the
quasiparticle picture.  Which alternative is realized depends on the behavior
of the interaction function $f$.
Before reaching the QCP, the standard ground-state FL distribution $n_F(p)$
applies because the necessary stability condition (NSC) for the Landau state
is satisfied.  This condition reads simply\cite{physrep,jetplett2011}
\beq
\delta E_0=\int\epsilon( p;n_F) \delta n(p)d\upsilon> 0.
\label{nesc}
\eeq
In words, the NSC is satisfied provided that for any admissible, particle-number
conserving variation $\delta n (p)$ from the standard Landau quasiparticle momentum
distribution $n_F(p)$, the change of the ground-state energy $E_0$ turns out to
be positive.  This is the case in naturally occurring Fermi liquids, such as
3D liquid $^3$He or low-density neutron matter, where one has $\epsilon(p)=v_F(p-p_F)$
with a positive value of the Fermi velocity $v_F=p_F/M^*$.  In this case,
the NSC is always met and the Fermi surface is singly connected, since the
sign of the variation $\delta n$ coincides with sign of the difference $p-p_F$
and hence with the sign of $\epsilon(p)$.

However, consideration of the full Lifshitz phase diagram anticipates situations
in which at a threshold value of some input parameter, such as the density or
magnetic field, there is a discrete change in the number of roots of
fundamental equation\cite{volrev}
\beq
\epsilon( p)=0.
\label{topeq}
\eeq
This change occurs, for example, at a critical density $\rho_{\infty}$ where
the Fermi velocity $v_F(\rho_{\infty})$ vanishes, or, equivalently, the effective
mass $M^*(\rho_{\infty})$ diverges---i.e., at the QCP.  Beyond this point, the number
of roots of Eq.~(\ref{topeq}) increases stepwise, implying that the Fermi surface becomes
multi-connected.  In other words, the minimum of $E(n)$ leaves the Landau
point $n_F(p)$ and moves to other {\it boundary points} of the manifold
$\{n(p)\}$ of admissible momentum distributions, at which $n(p)$ continues
to take only the two values 0 and 1.  In this situation FL theory still holds,
at least when the temperature remains sufficiently low.  This is the alternative
that applies specifically to a regular interaction function $f$.

A qualitatively different transformation appears when the coupling strength of
a regular interaction, here denoted by $\lambda$, is further increased to another
critical value $\lambda_c$, at which the minimum of $E(n)$ at a given particle number
eventually sinks into the {\it interior} of the manifold $\{n\}$.  The
true ground-state momentum distribution $n(p)$ within the extended Landau
quasiparticle picture then becomes a {\it continuous} function $n_*(p)$
of $p$ determined by Eq.~(\ref{varen}), the chemical potential $\mu$ being
identified as a Lagrange multiplier associated with particle number-conservation.
It will be convenient to refer to such distributions $n_*(p)$ as {\it non-idempotent
solutions} of the variational problem (\ref{varen}) on the manifold $\{n\}$,
meaning that they visit interior points $0 < n(p) < 1$ of the closed interval
$[0,1]$, whereas boundary solutions $n(p)$ have values restricted to 0 or 1
and hence are idempotent in the sense that $n^2 = n$.
 For a singular interaction, a nonidempotent solution appears already at the QCP.

To illuminate the nature and conceptual status of these novel NFL solutions,
we invoke a mathematical correspondence of the functional $E(n)$ with the
energy functional $E(\rho)$ of statistical physics as taught from textbooks.
If the interactions are weak, the latter functional attains its minimum value
at a density $\rho$ determined by the size of the vessel that contains
it, and which it fills uniformly.  In such cases, solutions of the
variational problem evidently describe {\it gases}.  On the other hand,
if the interactions between the particles are sufficiently strong,
there emerge nontrivial solutions of the variational condition
\beq
{\delta E(\rho)\over \delta \rho( r)}=\mu
\label{varrho}
\eeq
that describe {\it liquids}, whose density is practically
independent of boundary conditions.

\begin{figure}[t]
\includegraphics[width=0.8\linewidth,height=1.\linewidth]{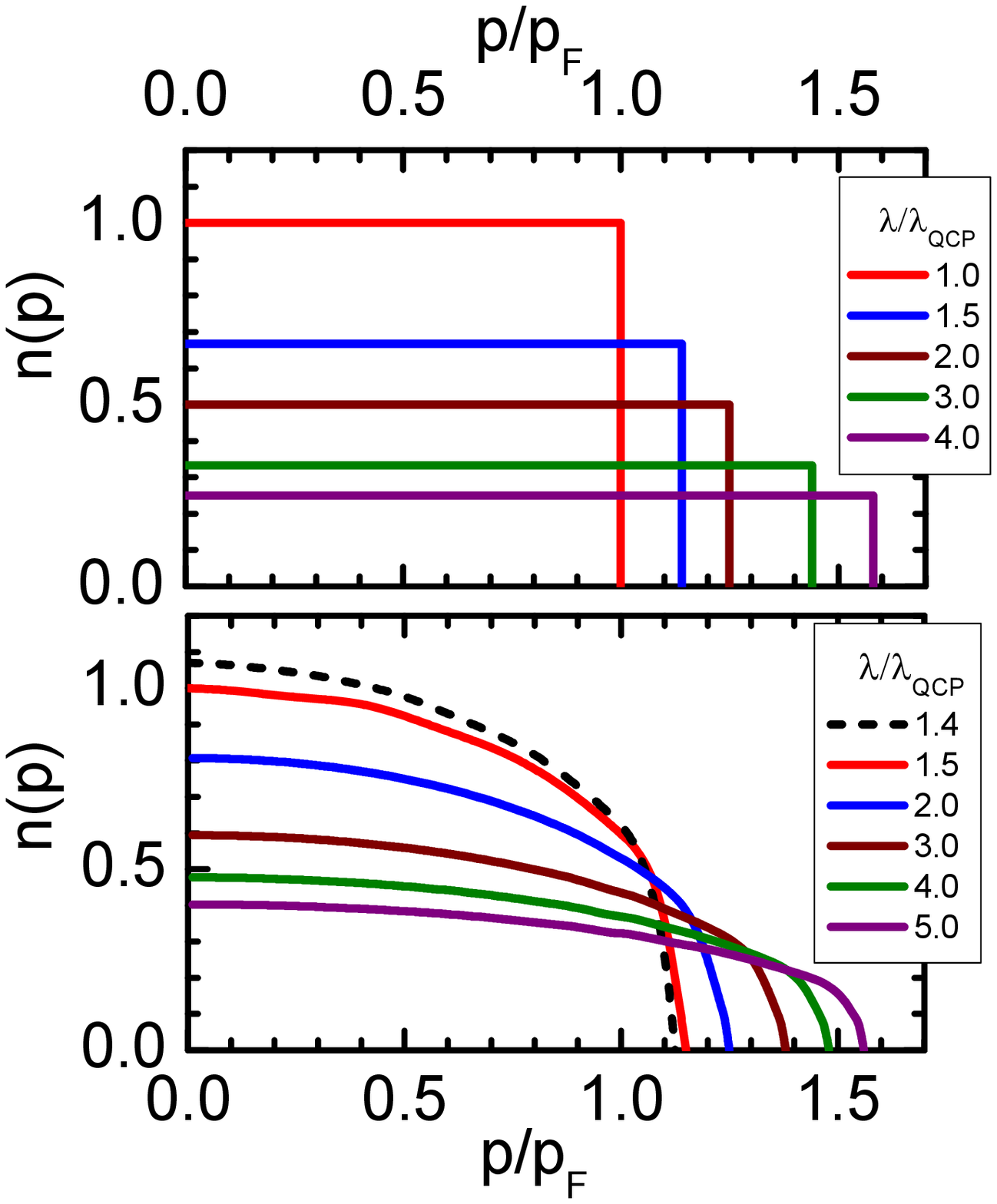}
\caption{
Occupation numbers $n(p)$ evaluated at different values of the ratio
$\lambda/\lambda_{QCP}$ for model (\ref{ksm}) [top panel], and
model (\ref{sfl}) [bottom panel].
}
\label{fig:occup}
\end{figure}

The energy functional $E[n]$ of our quantum many-body problem must have
two analogous types of solutions, with an essential difference: solutions
$n_*(p)$ of the variational condition (\ref{varen}) must satisfy the
condition $0\leq n(p)\leq 1$ imposed by fermion statistics.  This condition
cannot be met in weakly correlated Fermi systems, but it can be satisfied
in systems with sufficiently strong correlations.  We illustrate these points
by comparing the behavior of two models, one having a singular interaction
function\cite{ks} $f$ and the second having a regular interaction, smooth in
momentum space.\cite{prb2008}  Results based on the condition (\ref{varen})
are presented in the upper panel (singular interaction) and lower panel
(regular interaction) of Fig.~\ref{fig:occup}.

For both models, the energy functional takes the form
\beq
E(n)=\int {p^2 \over 2M} n({\bf p})\, d\upsilon
+{1\over 2}\int f({\bf p}_1-{\bf p}_2)n({\bf p}_1)n({\bf p}_2)
d\upsilon_1\,d\upsilon_2,
\label{ksm}
\eeq
the interaction function being given by
\beq
f({\bf p}_1-{\bf p}_2)={\lambda\over |{\bf p}_1-{\bf p}_2|}
\eeq
in the singular case (first model) and being represented by
\beq
f({\bf p}_1-{\bf p}_2)={\lambda\over ({\bf p}_1-{\bf p}_2)^2+\kappa^2}
\label{sfl}
\eeq
in the regular case (second model), with $\kappa = 0.07\,p_F$.

For the first model, the variational condition (\ref{varen}) reads
\beq
{p^2\over 2M}+\lambda\int
{n({\bf p}_1)\over |{\bf p}_1-{\bf p}|} d\upsilon_1  = \mu,
\label{ele}
\eeq
as is consistent with equation (\ref{lansp}).  For the second model, this
condition has the form
\beq
{p^2\over 2M}+\lambda\int
{n({\bf p}_1)\over ({\bf p}_1-{\bf p})^2 +\kappa^2} d\upsilon_1  = \mu.
\label{ese}
\eeq

The first of these equations is solved analytically by exploiting the aforementioned
mathematical correspondence between the functionals $E(n)$ and $E(\rho)$, where
$\rho(r)$ is the number density of a system of charged particles moving in an external
potential field and interacting with each other via Coulomb forces.  The result is simply
$n_*(p) =1/\lambda$ for $p < p_f$ and $n_*(p)=0$ otherwise, with the boundary
momentum $p_f$ being determined by particle-number conservation, i.e.,
$n_*p_f^3/3\pi^2=\rho$.  This qualifies as a legitimate non-idempotent solution of the
quantum problem under consideration if it obeys the Pauli exclusion principle,
which requires that the coupling constant $\lambda$ exceeds the critical
value $\lambda_c=\lambda_{\rm QCP}=1$ coincident with the value of $\lambda$
at which the FL effective mass diverges.  Evidently this solution does not
apply to weakly interacting systems characterized by $\lambda < \lambda_c$,
for which $n(p) = n_F(p)$ is the proper solution.

Turning to the second model, again there exists a non-idempotent solution of
Eq.~(\ref{ese}) for every $\lambda$ value.  However, as in the first model, this
solution fails to meet the Pauli principle at small coupling constants, implying that
in this case, the Fermi step $n_F(p)$ must be the ground-state momentum distribution
of the problem.  However, in contrast to the first model, when
$\lambda$ exceeds $\lambda_{QCP}$ (where the effective mass diverges and hence
the Fermi step $n_F(p)$ can no longer be the true ground-state quasiparticle
momentum distribution), the correct solution of the problem continues to
reside on the {\it boundary} of the manifold $\{n\}$, implying a multi-sheet Fermi
surface.  This is so because the non-idempotent solution of the variational condition
(\ref{varen}) still contradicts the Pauli restriction, and only when $\lambda$
attains a new critical value $\lambda_{\rm crit}=1.52\,\lambda_{QCP}$ where the
Pauli principle becomes immaterial, does the $n_*$ solution apply.
(For more detail, see Refs.~\onlinecite{zb,prb2008}).

\subsection{Insights from introduction of an external magnetic field}

Imposition of a magnetic field $H$ on systems without a spontaneous magnetic moment
splits the quasiparticle system into two subsystems, whose momentum distributions
$n_+(p)$ and $n_-(p)$ correspond to the $\pm 1/2$ projections onto the direction
of the magnetic field.   There is a consequent doubling of equilibrium condition
(\ref{varen}), implying
\beq
{\delta E(n_+,n_-)\over \delta n_+(p)}={\delta E(n_+,n_-)\over \delta n_-(p)}=\mu  .
\label{varhe}
\eeq
Let us explore this extension in the framework of a modified Nozi\`eres model\cite{noz}
in which both components $f$ and $g$ of the interaction function have the simplest
form,
$ f({\bf p}_1-{\bf p}_2)\propto g({\bf p}_1-{\bf p}_2)\propto\delta({\bf p}_1-{\bf p}_2)$,
such that the set of equations (\ref{varhe}) becomes algebraic.  At $T=0$, one has
\begin{eqnarray}
{1\over 2}\mu_BH+{p^2\over 2M}+{f+g\over 2}n_+(p)+{f-g\over 2}n_-(p)&=&\mu \  , \nonumber\\
-{1\over 2}\mu_BH+{p^2\over 2M}+{f-g\over 2} n_+(p)+{f+g\over 2}n_-(p)&=&\mu \  .
\label{seth}
\end{eqnarray}
In writing these equations we assume sign of the spin-spin component to be
repulsive.  Otherwise the system acquires a spontaneous magnetic moment,
and the corresponding analysis lies beyond the scope of this article.

A salient feature of this problem is revealed by subtracting the lower equation of
the set (\ref{seth}) from the upper one to obtain
\beq
0=\mu_BH+g(n_+(p)-n_-(p)).
\eeq
Since the right side of this equation is just the effective magnetic field acting
on the quasiparticle in the medium, we see that the magnetic field is completely screened
in the interior of the system, prompting an analogy with superconducting systems
(explained as fermion-pair condensates).
 The solution
\beq
n_{\pm}(p)={\mu-p^2/2M\over f}\mp{\mu_BH\over 2g}
\label{solh}
\eeq
of the set of equations (\ref{seth}) exists in the momentum interval
$p_i^{\pm}<p<p_f^{\pm} $ where the Pauli restrictions $0\leq n_{\pm}(p)\leq 1$
are met.  The expulsion of the external magnetic field remains in effect until
its magnitude becomes so large that all of the occupation numbers (\ref{solh})
reach limits enforced by the Pauli restrictions.
With this stipulation, the phenomenon of complete suppression of the magnetic
field acting on a quasiparticle, while derived here within a special model,
remains generally applicable---as demonstrated by numerical calculations
based on the set (\ref{seth}).  However, at finite temperatures complete
suppression occurs only in fields that are sufficiently strong, i.e., such
that $\mu_BH>T$; otherwise the screening effect disappears.\cite{yak}

\subsection{Fermion condensation and the Lifshitz phase diagram}

We now address more directly the properties of the NFL solutions
of Eq.~(\ref{varen}).  Recalling the definition (\ref{spec}) of the
quasiparticle energy, Eq.~(\ref{varen}) is recast economically as
\beq
\epsilon(p)=0 , \quad p\in \Omega  .
\label{efc}
\eeq
This means that the spectrum $\epsilon(p)$ of the strongly correlated Fermi system,
which coincides with the solution of the problem obtained within the refined
iterative procedure, must be
{\it completely flat} in some domain $\Omega$ of momentum space,
implying that the Fermi actually swells to become a volume in momentum space.

The set of states for which Eq.~(\ref{efc}) is satisfied has been called the
{\it fermion condensate} (FC).  The phase transition in which the Fermi surface
swells from a line to a surface in 2D, or from a surface to a volume in 3D
(discovered 20 years ago\cite{ks,vol,noz,physrep}) was originally called fermion
condensation---and more recently, swelling the Fermi surface or emergence of
{\it flat bands}.
The rearrangements of the standard Landau state that occur at and following the
QCP, namely (i) the generation of additional Fermi surfaces (or pockets) and
(ii) fermion condensation (or formation of a flat band) have been identified
as symmetry-preserving topological phase transitions (TPT's).\cite{prb2008,volrev}
In finite Fermi systems, the phenomenon analogous to fermion
condensation expresses itself as the {\it merging} of neighboring single-particle
levels.\cite{haochen}  Interest in flat bands was ignited
recently\cite{vol2011b,vol2011c,vol2011d,kopnin} in connection with a
related opportunity to raise the critical temperature $T_c$ for termination
of superconductivity of the electron systems in solids.

Another salient feature concomitant with flattening of the single-particle
spectrum is {\it partial occupation of single-particle states of given
spin}.  As already seen, in the presence of strong correlations, the momentum
distribution $n(p)$ is no longer restricted to the values 0 and 1, but may take
any value {\it within} the interval $[0,1]$.  Aided by the variational condition
(\ref{varen}), one can acertain what happens in this case when a quasiparticle
with momentum ${\bf p}\in \Omega$ is added to the system, again assumed to
be homogeneous.  In contrast to what occurs in a canonical Fermi liquid, addition
of just one quasiparticle now induces a rearrangement of the {\it whole} distribution
function $n_*(p)$ within the domain $\Omega$.  It follows that the system can
no longer be described as a gas of interacting quasiparticles, even though
the original Landau quasiparticle concept still applies.

In the intermediate region of the Lifshitz phase diagram, corresponding to
$\lambda_{QCP}<\lambda<\lambda_{FC}$ in the second example of Section.~V.A,
the Fermi surface still remains multi-connected.  A striking peculiarity of the segment
of the Lifshitz phase diagram just beyond the QCP, revealed in Ref.~\onlinecite{zb},
is the fast breeding of new small pockets of the Fermi surface as $\lambda$ increases
beyond $\lambda_{QCP}$.  The occurrence of such cascades is associated with the condition
$n(p)\leq 1$ enforced by the Pauli principle.  As we have seen earlier, solutions
of the equilibrium equation (\ref{varen}) always exist, but these solutions
violate the restriction $n(p)\leq 1$ until the coupling constant $\lambda$ attains
a critical value $\lambda_{FC}$. It is this restriction that triggers a cascade
of topological phase transitions in the segment of the Lifshitz phase diagram defined by
$\lambda_{QCP}<\lambda<\lambda_{FC}$.

We are now ready to comment on the generic structure of the $T=0$ Lifshitz phase
diagram of a homogeneous Fermi system.  This diagram contains a FL domain, where
correlations between particles are not so strong and the Fermi surface is
singly connected.  The quantum critical point separates this domain from a
region where the Fermi surface acquires a new small pocket.  In turn, this
region adjoins a stretch of the diagram in which the Fermi surface contains
two new pockets, and so on.  In all these states, the FL relation $n^2(p)=n(p)$
of idempotency is preserved.

The discrete generation of new pockets (or sheets) of the Fermi surface continues
up to $\lambda=\lambda_{FC}$.  At that point there emerges a new ground state
that solves the variational condition (\ref{varen}) with a non-idempotent
momentum distribution $n_*(p)$ satisfying the Pauli restriction $0\leq n_*(p)\leq 1$
everywhere.  This state wins the energetic contest with any state
having a discretely multi-connected Fermi surface.  It is this exceptional
state that possesses a finite entropy excess $S_*$, in apparent contradiction
to the Nernst theorem.

So far we have restricted attention to the case of zero temperature.  At
finite $T$ the degeneracy of the FC is lifted as an integral feature of the
physics of strongly correlated Fermi systems,\cite{noz} and the spectrum
$\epsilon(p,T)$ acquires a finite slope with respect to $T$ in the domain
$\Omega$.  To determine this slope, Eq.~(\ref{dist}) can be inverted to obtain
\beq
\epsilon(p)= T\ln {1-n(p)\over n(p)}.
\label{eslope}
\eeq
Since a minute temperature elevation cannot affect the FC momentum distribution
$n_*(p)$, it can be inserted for $n(p)$ on the right-hand side of this equation,
yielding the dispersion
\beq
\epsilon(p,T)= T\ln {1-n_*(p)\over n_*(p)} , \quad p\in \Omega,
\label{spte}
\eeq
of the FC spectrum at $T\to 0$.\cite{noz}
The degeneracy of the FC spectrum is thereby removed, the FC dispersion now
being proportional to $T$.

A remarkable feature of the enlarged picture of the phase diagram,
considered above at $T>0$ in the context of the FC solution, is that
the behavior (\ref{spte}) prevails not only at $\lambda\simeq \lambda_c$,
but already makes its appearance at $\lambda\geq \lambda_{QCP}$, i.e.\
just beyond the quantum critical point.\cite{prb2008}  To make this
point more quantitative, consider that in a rearranged ground state
at $T=0$ having a multi-connected Fermi surface, the spectrum $\epsilon(p,T=0)$
shows drastically different behavior in different regions of the momentum
variable $p$.  It varies smoothly with $p$ in regions away from the pockets
of the Fermi surface, but oscillates very rapidly where the pockets reside,
since the number of zeros of $\epsilon(p,T=0)$ must equal the number of
pockets.  The departure of $|\epsilon(p,T=0)|$ from zero in the latter domain,
denoted below by $T_m$, emerges as a new energy scale of the problem,
around which a ``melting'' of the structure associated with the discretely
multi-connected Fermi surface takes place.  Indeed, as seen from the basic
Landau formula (\ref{dist}), the distribution $n(p,T)$ remains almost
unchanged from $n(p,T=0)$ as long as $T<T_m$.  However, as the temperature
$T$ increases, kinks in $n(p,T)$ versus $p$ are gradually smeared, and
at $T>T_m$, the function $n(p,T)$ becomes continuous and almost $T$-independent
in the momentum region where its abrupt variations occur at $T=T_m$.
Accordingly, Eq.~(\ref{spte}) comes into play and the dispersion of the 
single-particle spectrum $\epsilon(p)$ becomes proportional to $T$.\cite{prb2008}
{\it In effect, at $T>T_m$ the temperature evolution of $\epsilon(p,T)$
and $n(p,T)$ becomes universal, coinciding with their evolution in the
state having a FC}.

\begin{figure}[t]
\includegraphics[width=0.8\linewidth,height=0.65\linewidth]{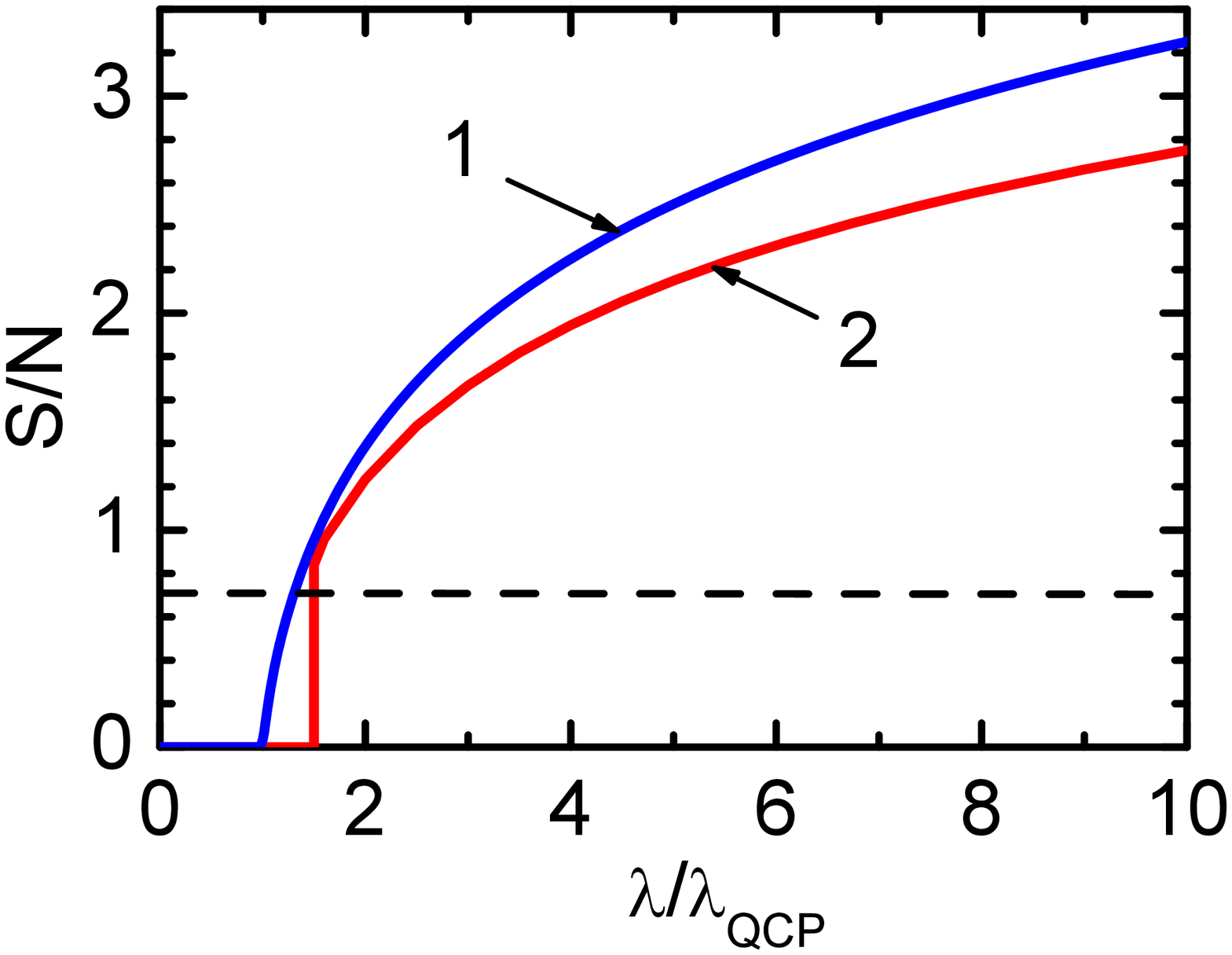}
\caption{
Entropy per particle $S/N$ versus the ratio $\lambda/\lambda_{QCP}$ for
the models (\ref{ksm}) and (\ref{sfl}), which are identified by ``1'' and
``2,'' respectively. The value of $\ln 2$, inherent in a typical chaotic
solution $n=1/2$, is shown by horizontal dashed line.
}
\label{fig:entropy}
\end{figure}

Since the occupation numbers $n_*$ are $T$-independent in systems having a FC,
the entropy $S(T\to 0)$ determined by Eq.~(\ref{entr}) has a nonzero value,
as already foretold by the refined iteration/time average procedure for solving
the basic equation (\ref{lansp}).  To illustrate this important property,
Fig.~(\ref{fig:entropy}) presents results of numerical calculations of the
entropy $S(T=0)$ for the model energy functionals (\ref{ksm}) and (\ref{sfl})
that contradict the Nernst theorem, which requires $S(T\to 0) =0$.
The contradiction is not evaded by going to finite $T$, even though the degeneracy
of the single-particle spectrum is lifted.  Elimination of the degeneracy
does not lead to recovery of the basic FL relation $n^2(p)=n(p)$, because
the quasiparticle momentum distribution $n(p)$ remains essentially
the same as it was at $T=0$.

%

\section{Resolution of the entropy paradox}

The first possible route to removal of the entropy excess $S_*$, which is
appropriate to the part of the Lifshitz phase diagram where the $T=0$
Fermi surface is multi-connected (i.e., $\lambda_{QCP}<\lambda<\lambda_{FC}$),
has been already analyzed above, although not identified as such.  The
entropy $S_*$ is released by means of a {\it crossover} at $T\simeq T_m$ from
the FC state to a state with a multi-connected Fermi surface (or a Fermi
surface decorated with pockets).  It is likely that such topological phase
transitions have already been observed experimentally\cite{oeschler,hossain}
in the heavy-fermion metals CeCoIn$_5$ and YbIr$_2$Si$_2$.

Alas, this prescription for shedding the excess entropy is unavailable in the
strong-coupling limit, when all the quasiparticles reside in the FC region.  In
this case, release of the entropy excess as $T\to 0$ can occur in a different
way.  At this point it is important to recognize that in deriving the basic
formulas (\ref{part}), (\ref{dist}), and (\ref{lansp}), we tacitly assumed that
any phase transition involved in entropy release take places {\it without}
breaking any symmetry inherent in the Landau ground state.  With this in mind,
resolution of the entropy paradox remains an open problem from the theoretical
standpoint, though it is surely decided in nature from among a number of possible
routes for draining the entropy excess.  Different phase transitions exist as
candidates even among those that preserve homogeneity of the system.  Of
these, it is instructive to focus attention on superconducting phase
transitions.  This route was already considered in this context in the
first article\cite{ks} devoted to fermion condensation, because the
preservation of homogeneity and isotropy in the superconducting ground
state and the attendant violation of gauge invariance.  By virtue of Cooper
pairing, the spectrum $E(p)$ of single-particle excitations of the superconducting
ground state has a gap $\Delta$.  It is this gap that ensures a vanishing
entropy at $T=0$ in agreement with the Nernst theorem, as seen in the
BCS relation
\beq
S(T\to 0)\propto e^{-\Delta(0)/T}.
\eeq
Thus, one reliable way to eliminate $S_*$ in systems with the well-developed FC
is associated with Cooper pairing.  The superfluid/superconducting phase transition
takes place provided the particle-particle interaction is attractive.  In
liquid $^3$He, this interaction is certain to be attractive in channels with
large orbital angular momenta $L$.\cite{pit}  At any rate, in 3D liquid $^3$He,
pairing takes place in the triplet spin channel with $L=1$.  On the other hand,
in electron systems of solids, pairing typically occurs in singlet channels.

Within the framework of BCS theory, the critical temperature $T_c$ of the
superconducting phase transition in the singlet channel is determined from
the BCS gap equation
\beq
\Delta(p,T)=\int {\cal V}(p,p_1){ \tanh \left(E(p_1)/2T\right) \over 2E(p_1)}
\Delta(p_1,T) d\upsilon_1  ,
\label{bcs}
\eeq
where ${\cal V}$ denotes the effective interaction between quasiparticles in the
corresponding pairing channel, while $E(p)=\left[\epsilon^2(p)+\Delta^2_L\right]^{1/2}$.
As usual, the momentum dependence of the gap function is neglected.
At $T\to T_c$ the gap $\Delta_L$ vanishes, and we are left with a linear integral
equation for the single unknown parameter, the critical temperature $T_c$.
In the strong-coupling limit, where at $T=0$ all the quasiparticles belong to the FC,
the critical temperature $T_c$ is expressed in terms of the FC density with the aid
of the identity
\beq
{{\tanh \left( \epsilon(p)/2T_c \right) }\over {\epsilon(p)}}
={1-2n_*(p)\over 2T_c\ln \left[(1-n_*(p))/n_*(p)\right]}.
\eeq
Straightforward algebra leads to
\beq
T_c =\int {\cal V}(p,p_1) {1-2n_*(p_1)\over 2\ln
\left[(1-n_*(p))/n_*(p)\right]} d\upsilon_1.
\label{bcs1}
\eeq
In the strong-coupling regime, where the FC density coincides with the density $\rho$,
the critical temperature $T_c$ may then be estimated from this formula as
\beq
T_c\simeq {\cal V}\rho,
\label{tcfc}
\eeq
where ${\cal V}$ is a constant measuring an averaged pairing interaction. Since
$\Delta(T=0)\simeq 1.76T_c$ in BCS theory, we infer that the critical temperature
associated with release of the excess entropy is close to $T_c$, which turns
out to be of order of the Fermi energy, except for the case of the extremely
small pairing constant.

This situation applies in heavy-fermion metals, which possess electron systems
having narrow $4f$-bands, where NFL behavior---including the presence of the entropy
excess---is well documented.  However, in these systems the FC density is rather small
and then, as discussed earlier, the route to recovery of the Nernst theorem
via the mechanism of singlet BCS pairing is not unique.  Furthermore, this route
can be blocked by imposition of a modest external magnetic field, sufficient to
terminate superconductivity.  As we have seen, imposition of an external
magnetic field affects the FC only weakly, while the effect of the field
on superconductivity, ultimately its destruction, is far more pronounced.
Without attending to cumbersome formulas, we know that the critical
magnetic field $H_c$ sufficient for termination of superconductivity at zero
temperature is proportional to the gap value $\Delta_L(0)\propto T_c$.
Thus we can expect that imposition of modest magnetic fields $H \simeq H_c$ on
a superconducting system that hosts a FC in its normal state, i.e.\ at $T>T_c$,
causes re-entry of the system into such non-superconducting FC-bearing
states at temperatures lower than $T_c$, thus moving the critical
temperature for the onset of the Nernst regime toward zero.

\begin{figure}[t]
\includegraphics[width=1.2\linewidth,height=1.5\linewidth]{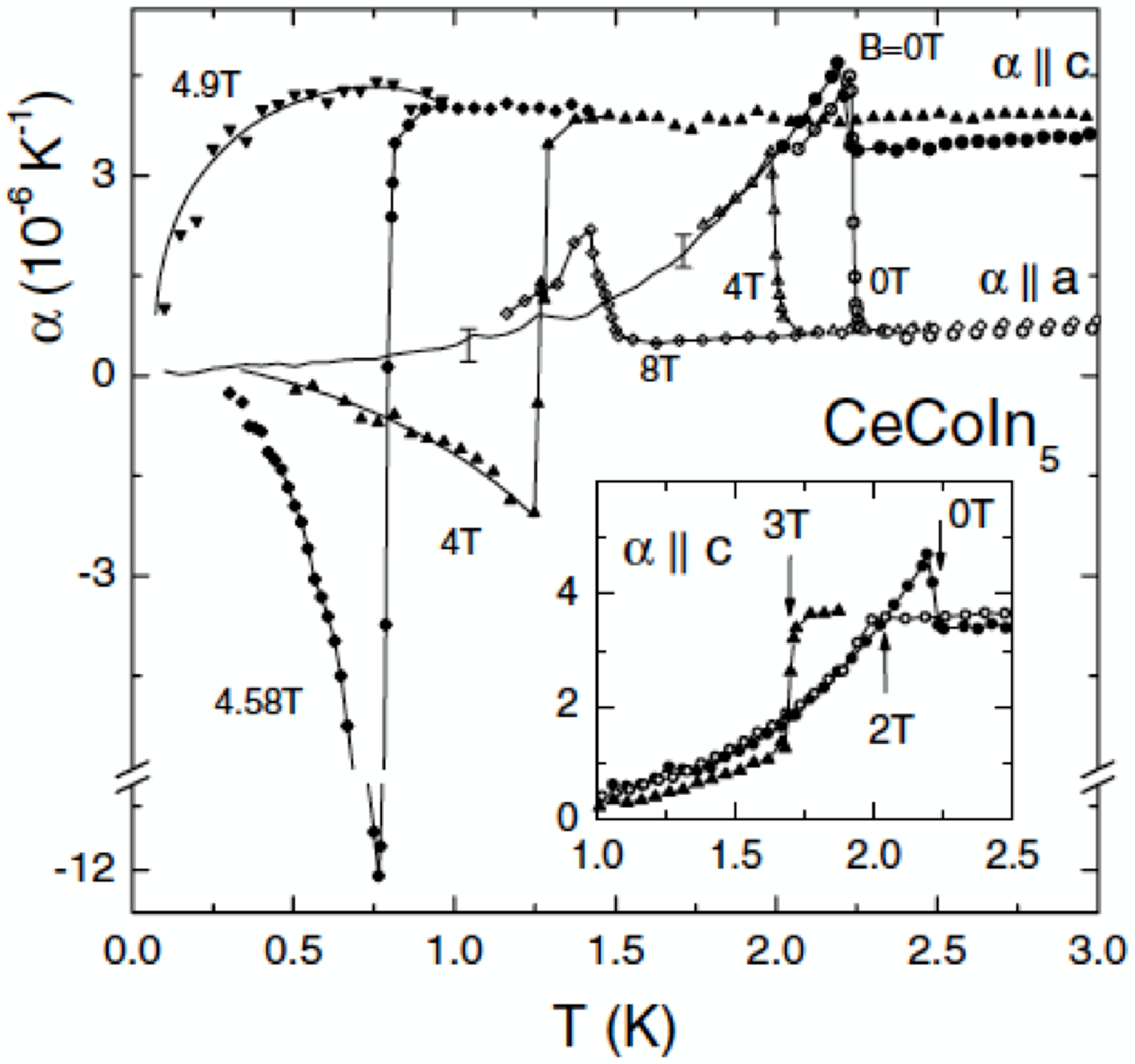}
\caption{ Low-temperature thermal expansion coefficient of the heavy-fermion
metal CeCoIn$_5$ at different magnetic fields $B$ (from Ref.~\onlinecite{oeschler},
with permission).
}
\label{fig:cecointherm}
\end{figure}

Is there any experimental evidence supporting the validity of this proposal,
which predicts quite unexpected behavior of strongly correlated Fermi
systems at low temperatures?  Analysis of experimental data\cite{oeschler}
on the thermal expansion coefficient $\alpha$ of the heavy-fermion metal
CeCoIn$_5$ reveals just such behavior in this material, which has the extremely
low value $T_c = 2.3$ K for the critical temperature of its superconducting
phase transition, compared with the characteristic energy scale of order
$10^3$ K for the electron system of heavy-fermion metals.  Data taken from
Ref.~\onlinecite{oeschler} are displayed in Fig.~\ref{fig:cecointherm}.
We observe first that under decreasing temperature with no external magnetic
field present, $\alpha$ does in fact undergo a jump at critical temperature
$T_c$, affirming that one is dealing with a superconducting second-order phase
transition, with $\alpha(T)$ subsequently dropping off to 0 at $T\to 0$,
all as in the standard BCS regime.  Second, imposition of a magnetic
field $H$ ($B$ in the figure) to 4 T and then 4.58 T shifts the jump
of $\alpha(T)$ toward $T=0$, in accordance with the BCS trajectory of the
line $T_c(H)$.

However, interpretation of further details of the thermal evolution of
the coefficient $\alpha(T)$ presents a special challenge: at $T>T_c=2.3 K$
the value of $\alpha(H=0)$ is $T-$independent and greatly enhanced
compared with values typical for ordinary metals.  The behavior of $\alpha(H)$
under variation of an imposed magnetic field defies explanation within standard
theoretical
approaches.
As $H$ increases and $T_c(H)$ correspondingly declines, the value of the
jump in $\alpha$ at $T=T_c$ grows rapidly, ultimately leading to recovery
of the anomalous values of $\alpha$ that are found at $T>T_c$ in the
absence of the magnetic field (see Fig.~\ref{fig:cecointherm}).
We suggest that this perplexing NFL behavior can be explained in a natural
way.  Let us assume that the behaviors in $H$ and $T$ of the thermal
expansion coefficient and the entropy are alike.  Further, suppose the
system carries an entropy excess due to the presence of a small portion
of FC---which, as we have seen, is quite insensitive to the magnitude of
$H$ but highly sensitive to the presence of pairing correlations.  It then
follows that at $H>H_c$, the normal-state thermal expansion coefficient
begins to decline only at temperatures lower than 1 K, furnishing evidence
for the suppression of the entropy excess in harmony with the quest for
satisfaction of the Nernst theorem.  We suspect that this decline, which
occurs without any jumps, can be explained as a crossover
(or other transition)
from the state with a FC to a state with a multi-connected Fermi surface,
as discussed above.

Based on these experimental facts and theoretical considerations, we predict that
analogous anomalies should be observed in measurements of the low-temperature
thermal expansion coefficient of another heavy-fermion compound, namely
the $P$-type of the heavy-fermion metal YbIr$_2$Si$_2$, whose entropy
$S(T)/N $ attains values around $\ln2$ already at extremely low
temperatures $T\simeq 1 K$.\cite{hossain}

In systems with a well-developed FC, there exist other mechanisms that operate
as effectively as singlet pairing correlations. Among them are triplet pairing
correlations and the BCS-BEC crossover. However, as a preliminary analysis shows,
the corresponding critical temperatures may be even lower than that associated
with singlet pairing.

\section{Conclusion}
We have explored the entropy paradox encountered in a many-fermion system when
the correlations between particles become large enough to produce a
rearrangement of the standard Landau state.  The paradox arises from the
fact that the system retains a large, temperature-independent entropy
excess down to extremely low temperatures, behavior that is in apparent
contradiction to the Nernst theorem.

We have demonstrated that this unorthodox behavior can be explained within the
general framework of the Landau quasiparticle theory.  We find that it stems
from a swelling of the Fermi surface that occurs when the correlation strength
reaches a critical value. Equivalently, there is a flattening of the single-particle
spectrum near the Fermi surface, with concomitant smearing of the ground-state
quasiparticle momentum distribution.  This phenomenon has been examined along
two lines:
\begin{itemize}
\item[(i)]
Analysis of a Poincar\'e mapping corresponding to the fundamental Landau equation,
which connects the single-particle spectrum with the quasiparticle momentum
distribution through the amplitude of the quasiparticle interaction, and
\item[(ii)]
Solution of the variational condition for the minimum of the ground state
energy with a subsidiary restriction to ensure particle-number
conservation.
\end{itemize}
We have investigated routes for releasing the entropy excess as the temperature
falls to zero, thereby avoiding violation of the Nernst theorem.  Focusing
on scenarios in which the system remains homogeneous, we have found that
phase transitions associated with Cooper pairing are most effective in
suppression of the entropy excess.  We have shown that this conclusion is
consistent with experimental data on the low-temperature behavior of the
thermal expansion coefficient of the heavy-fermion metal CeCoIn$_5$.

This research was supported by the McDonnell Center for the Space Sciences, 
by Grant No.~NS-215.2012.2 from the Russian Ministry of Education and Science, 
and by Grant No.~12-02-00804 from the Russian Foundation for Basic Research.

\end{document}